\documentclass[sigconf]{acmart}

\usepackage{algorithm} 
\usepackage{algpseudocode} 
\usepackage{listings}
\usepackage[inline]{enumitem}
\usepackage{array}
\usepackage{tabularx}
\usepackage{subcaption}

\AtBeginDocument{%
  \providecommand\BibTeX{{%
    \normalfont B\kern-0.5em{\scshape i\kern-0.25em b}\kern-0.8em\TeX}}}

\copyrightyear{2021}
\acmYear{2021}
\setcopyright{acmlicensed}
\acmConference[WSDM '21] {Proceedings of the Fourteenth ACM International Conference on Web Search and Data Mining}{March 8--12, 2021}{Virtual Event, Israel}
\acmBooktitle{Proceedings of the Fourteenth ACM International Conference on Web Search and Data Mining (WSDM '21), March 8--12, 2021, Virtual Event, Israel}
\acmPrice{15.00}
\acmDOI{10.1145/3437963.3441812}
\acmISBN{978-1-4503-8297-7/21/03}
\begin{document}
\fancyhead{}

\title{Mining the Stars: Learning Quality Ratings with User-facing Explanations for Vacation Rentals}
\author{Anastasiia Kornilova}
\email{anastasiia.kornilova@booking.com}
\affiliation{%
  \institution{Booking.com}
  \city{Amsterdam}
  \country{The Netherlands}
}

\author{Lucas Bernardi}
\email{lucas.bernardi@booking.com}
\affiliation{%
  \institution{Booking.com}
  \city{Amsterdam}
  \country{The Netherlands}
}


\begin{abstract}
Online Travel Platforms are virtual two-sided marketplaces where guests search for accommodations and accommodation providers list their properties such as hotels and vacation rentals. The large majority of hotels are rated by official institutions with a number of stars indicating the quality of service they provide. It is a simple and effective mechanism that contributes to match supply with demand by helping guests to find options meeting their criteria and accommodation suppliers to market their product to the right segment directly impacting the number of transactions on the platform. Unfortunately, no similar rating system exists for the large majority of vacation rentals, making it difficult for guests to search and compare options and hard for vacation rentals suppliers to market their product effectively. In this work we describe a machine learned quality rating system for vacation rentals. The problem is challenging, mainly due to explainability requirements and the lack of ground truth. We present techniques to address these challenges and empirical evidence of their efficacy. Our system was successfully deployed and validated through Online Controlled Experiments performed in Booking.com, a large Online Travel Platform, and running for more than one year, impacting more than a million accommodations and millions of guests.
\end{abstract}



\maketitle
\section{Introduction}
\begin{figure*}
\includegraphics[width=0.7\linewidth]{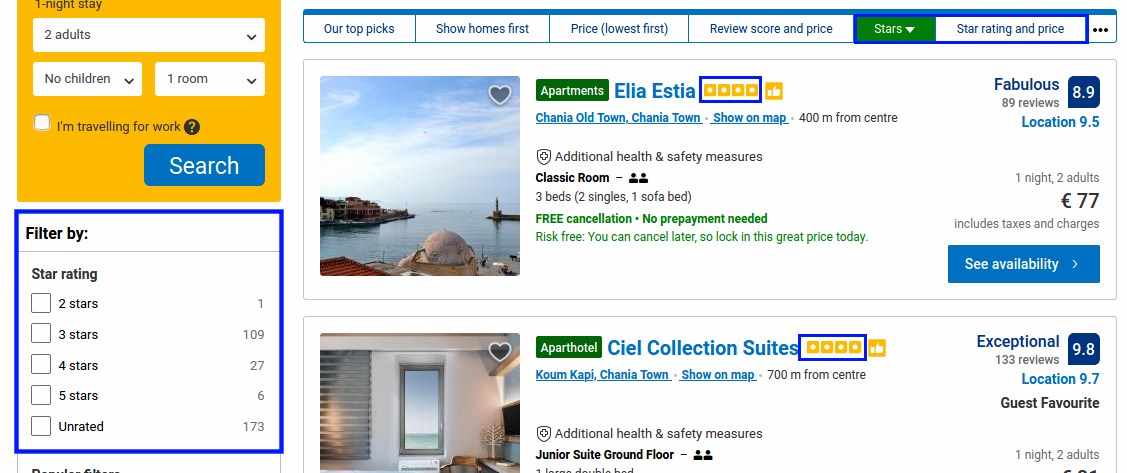}
\caption{Booking.com Search Results Page. In blue, highlighted tools relying on Quality Rating}
\label{fig:tiles}
\end{figure*}

Typical vacation rentals (VRs) are individually owned apartments or houses offering self-catering hospitality services on a temporary basis. They include a wide spectrum of property types such as professionally managed complexes, farm stays, apart-hotels, bed and breakfasts, etc. offering a broad variety of options alternative to the accommodation service provided by hotels. During the last years, VRs became very popular on Online Travel Platforms. One characteristic that distinguishes them from hotels is that the large majority lacks a star rating. Official institutions classify hotels in the well known 1-5 stars rating scale. This is a globally established system that customers know and understand, helping both the demand and the supply by creating realistic expectations about the quality of the service. It is also a very useful tool to navigate a large supply of accommodations alternatives through filtering and comparisons, which helps to better match demand with supply \cite{bernardi2019150} (see Figure \ref{fig:tiles}). Booking.com is an Online Travel Platform that offers both hotels and vacation rentals, which implies that when users apply star rating filters, the large majority of vacation rentals are immediately removed from the results list, which puts VRs at a clear disadvantage compared to hotels and hides potentially relevant options from the guests. This context motivates the need for quality ratings for vacation rentals that are comparable to hotels stars. One approach to consider is to classify VRs by expert assessment, but this is not a scalable solution since the experts need to actually visit each property listed in the platform. Remote classification suffers from high subjectivity and would produce ratings not comparable to hotels stars. In view of this, an automated VR rating process becomes an appealing solution. Accommodation quality assessment has many challenges, as described for the hotels case in \cite{vine1981hotel}. In our specific case, we focus on automated explainable vacation rental quality rating, which poses the following extra challenges:
\begin{itemize}
    \item \textit{Lack of Ground Truth}: The amount of officially rated vacation rentals is very small. This is discussed in Section \ref{sec:solution}.
    \item \textit{User-facing Explanations}: as the system is replacing a typically human task, explanations are critical to generate trust with the customers. This poses many challenges discussed in \cite{arrieta2020explainable} and \cite{gunning2017explainable}. Furthermore, explanations have business purposes like helping property managers maintain and improve the quality of their vacation rental quality. This is discussed in Section \ref{sec:modeling}.
    \item \textit{Hotels compatibility}: since hotels and vacation rentals are listed on the same platform, we want to make sure they are comparable. Specifically, the automated quality rating system must mimic the human task of visiting accommodation and assessing the provided quality. This is discussed in Section \ref{sec:solution}.
    \item \textit{User generated content}: the input for a specific rating is a description of the property generated by (typically non-professional) property managers. In some cases, these descriptions are incomplete and contain mistakes, making both labeling and explanations even more challenging. This is discussed in Section \ref{sec:mono}.
\end{itemize}
All of these challenges are addressed by our solution. Our main contributions are:
\begin{itemize}
    \item The description of a machine learning system capable of producing global and explainable vacation rental quality ratings
    \item Comparison of methods and techniques to address the mentioned issues
    \item A set of large scale online controlled experiments that independently show the effectiveness and business impact of: 
    \begin{itemize}
        \item Machine Learned generated VR Quality Ratings on both guests and property manager sides
        \item Explanations to property managers
        \item Suggestions for property managers to improve the rating of their property
    \end{itemize}
\end{itemize}
To the best of our knowledge, no prior work focuses on automated accommodation quality ratings. The closest work we aware of is related to predicting guest ratings in hotels \cite{leal2017prediction}, but this is very different from our setting since guest ratings are not comparable to hotels star ratings because they are based on guests experiences as opposed to expert assessment, and more importantly, they depend on the subjective expectations of each guest introducing too much variance in the rating distribution for one property. The paper is organized as follows: Section \ref{sec:problem} formalizes the problem, Section \ref{sec:solution} describes our approach to generate labeled data, Section \ref{sec:modeling} discusses our explanations aware modeling approaches, Section \ref{sec:explainability} dives in to our method to explain VR ratings, Section \ref{sec:suggestions} describes an actionable advice generation process, Section \ref{sec:experiments} presents online controlled experiments conducted in Booking.com, a leading OTP with millions of daily users and vacation rentals and Section \ref{sec:conclusion} presents our conclusions.

\section{Problem Definition}
\label{sec:problem}
Our objective is to construct a system capable of producing a quality rating given a description of a vacation rental property. The property description is a set of attributes such as facilities, amenities, size, number of rooms, etc. and the rating is an integer ranging from 1 to 5. The model must be capable of explaining the assigned ratings, more specifically, it must be able to explain what separates a given rating from worse ratings and to suggest what need to be added to reach the next level. It must be global, which means that it must be able to rate all VR property types in all countries ($\sim$200). The rating system must be compatible with the hotels star rating system, which means that it should be as close as possible to an objective process where an expert physically visits and assesses the property. Finally, since property descriptions and properties themselves are updated with new facilities and services, the model must be able to update ratings, explanations and suggestions as soon as new property details are available.

\section{Collaborative Labeling}
\label{sec:solution}
Defining the problem as a mapping from VR descriptions to ratings makes Supervised Learning a natural approach to solve it, but unfortunately, we don’t have enough labeled VRs. However, we have the following data sets at our disposal:
\begin{itemize}[topsep=2pt]
    \item Rated hotels: Global set of hotels with their star ratings (more than 500000)
    \item Rated vacation rentals: vacation rentals with official ratings from one specific country and VR type (about 40000)
    \item Unrated vacation rentals: Global set of unrated vacation rentals (more than 2 million) 
\end{itemize}
The set of rated hotels is large enough to train and test standard supervised learning algorithms. The set of rated vacation rentals is only suitable for validation since it is small and only from one country and for one VR type. In this section we describe two labeling approaches and compare them using the rated vacation rentals and rated hotels sets. Since the ratings are ordered and their distribution is far from uniform (see Table \ref{tab:classdist}) we use the \textit{macro-averaged Mean Average Error}\cite{baccianella2009evaluation} (MAMAE) which computes the Mean Average Error per class and averages over the classes:
\begin{equation}
  MAMAE = \frac{1}{c}\sum_{j=1}^{c}\frac{1}{|T_j|}\sum_{x\in T_j } |\hat{y}(x) - j |
  \label{eq:mamae}
\end{equation} where $c$ is the number of classes (5 in our case), $T_j$ is the set of instances with true class $j$ and $\hat{y}(x)$ is the predicted class for property description $x$. We also report weighted F1, a typical classification metric that ignores class order. In order to benchmark these labeling schemes, we trained multinomial classifiers using gradient boosting \cite{friedman2002stochastic}, specifically Gradient Boosted Trees (GBT) with all the available features (about 400 including facilities/amenities, size, number of rooms, services, etc, which due to commercial sensitivity cannot be disclosed).
\begin{figure*}
\includegraphics[width=0.8\linewidth]{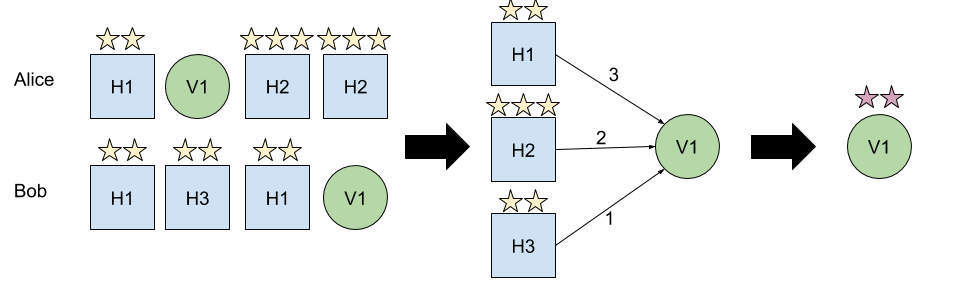}
\caption{Collaborative Labeling Process}
\label{fig:labelprop}
\end{figure*}

\begin{table}
\caption{Rating distributions of different data sets}
\small
\centering
\begin{tabular}{@{}lrrrrr@{}}
\toprule 
 & Rated Hotels & Officially Rated VRs  & Collaborative Labels VRs\\ \midrule
Class 1 & 5\%&0.5\%&0.2\%\\ 
Class 2 & 18\%&3\% &2.7\% \\ 
Class 3 & 45\%&80\%&66.9\% \\ 
Class 4 & 24\%&16\% &29\%\\ 
Class 5 &7\%&0.5\% &1.2\%\\ \bottomrule 
\end{tabular}
\label{tab:classdist}
\end{table}

\par The first approach consists of training a model on hotels using the star rating as a label. This approach works well for hotels (MAMAE 0.411), however, when evaluated on the rated vacation rentals set the performance is poor (MAMAE 1.01 on the full rated vacation rentals set). This is expected since the average hotel room is different from a vacation rental, typical VRs are equipped for self-catering facilities (kitchen, dishwasher, washing machine, etc.) which are not present in most hotels. This makes it very hard for a model to generalize from hotels to vacation rentals. 
\par In a second approach we apply a technique inspired by Label Propagation \cite{zhur2002learning} where we propagate hotel ratings to unrated vacation rentals. We construct a graph where vertices are unlabeled vacation rentals and labeled hotels. Each edge can only connect one hotel vertex $h$ with one vacation rental vertex $v$, and it is weighted by the number of stays in $h$ made by all guests who also stayed in $v$. We then construct a distribution over the star ratings based on the weights of all the edges of $v$.  Finally, the label is the mode of such distribution (see Figure \ref{fig:labelprop}). The main underlying assumption is that guests choose hotels and vacation rentals with similar quality of service. For example, if a user stayed at six different hotels, and most of them are 4-stars, if she then stays in a vacation rental $v$, we expect $v$ to be comparable to a 4-stars hotel. Effectively, user data is telling us how to transfer hotel star ratings to vacation rentals, hence, we name this technique \textit{Collaborative Labeling}. We validate the assumption by using the collaborative labels as predictions of the known hotel star ratings. We found performance comparable to training a model with star ratings indicating that the collaborative labels contain information about the true known hotel stars.  Furthermore, we trained a model with hotels data using collaborative labels and compared against training with the true ratings and found very similar performance indicating that training a model with collaborative labels is almost as good as using the true labels. Finally, we used the collaborative labels as predictions of vacation rentals ratings and evaluated with the small set of vacation rentals for which we do have an official rating. Performance is good, indicating that the collaborative labels also contain a lot of information about the true ratings of vacation rentals. All results are summarized in Table \ref{tab:labels}, showing that Collaborative Labeling is a sound technique to generate ground truth to train supervised machine learning models. 

\begin{table}
\caption{Performance of different Labeling Schemes}
\centering
\small
\begin{tabularx}{\linewidth} { 
  >{\raggedright\arraybackslash\hsize=0.45\hsize}X 
  >{\raggedleft\arraybackslash\hsize=0.16\hsize}X 
  >{\raggedleft\arraybackslash\hsize=0.21\hsize}X
  >{\raggedleft\arraybackslash\hsize=0.18\hsize}X}
\toprule
  & Trained on True Stars &Col. Labels as Predictions& Trained on Col. Labels\\ \midrule
MAMAE Hotels Val. Set & 0.411&0.525&0.588\\ 
MAMAE Labeled VRs & 1.01 &0.84&0.962\\ \midrule
Weighted F1 Hotels Val. Set&0.810&0.822&0.726\\ 
Weighted F1 Labeled VRs&0.662 &0.883&0.735\\ 
\bottomrule 
\end{tabularx}
\label{tab:labels}
\end{table}

\section{Explanation Aware Modeling}
\label{sec:modeling}
Through Collaborative Labeling we obtained a set containing more than 1 million labeled Vacation Rentals. We use this data to apply supervised learning techniques and rate all the remaining VRs, as well as new VRs, as they become part of the platform. But due to the explainability requirements, we have to make sure our models are capable of producing robust and consistent explanations, able to determine which characteristics are the main drivers of a specific rating \textit{with respect to the adjacent ratings} (as opposed to all other ratings). In other words, we want to explain to partners what makes a 3-stars property stand out from 2-stars properties, other 3-stars properties, and what is missing to reach 4-stars. Furthermore, which type of model is used to label properties has strong implications on the algorithms used to generate explanations like computation requirements and explanations semantics. We consider all these aspects while solving the prediction problem in order to guarantee accurate, scalable, and explainable models satisfying the established requirements. 
\subsection{Baselines}
\label{sec:baselines}
\par As a trivial baseline, we consider the most frequent rating (mode-classifier), which has about 75\% accuracy. Linear regression,  as a white-box model, is a natural approach to get an interpretable model \cite{tibshirani1996regression}, but we saw poor results (worse than the  mode-classifier). Although linear regression is able to capture the order in the classes, it predicts the expected value, which needs to be discretized to match the possible labels. Such discretization process is non-trivial, we experimented with various thresholding techniques, but performance was always below the baseline level.
\par Another natural approach is Ordinal Regression, where ratings are still considered discrete, but ordered. We applied Logistic Ordinal Regression\cite{harrell2015regression} which separates classes with parallel decision boundaries and it is also straightforward to explain \cite{bender1997ordinal}. We did see an improvement on MAMAE, indicating that class ordering is helping, but still worse than baseline on F1 and Accuracy. We hypothesize that both Linear and Ordinal Logistic Regression struggle to find linear decision boundaries in the sparse and mostly binary feature space. Therefore, with the aim of learning non linear decision boundaries, we turned to Multinomial Classification with Gradient Boosted Trees (GBT), for which efficient explanation generation algorithms exist \cite{lundberg2020local2global}, which showed much better performance on all Accuracy, F1, and MAMAE metrics, except for MAMAE on the Labeled VR Set, which suggests there is room for improvement by introducing ordered labels. Results are summarized in Table \ref{tab:ordmult}.
\begin{table}
\caption{Baselines performance}
\small
\begin{center}
\noindent
\begin{tabularx}{\linewidth}{ 
  >{\raggedright\arraybackslash\hsize=0.49\hsize}X
  >{\raggedleft\arraybackslash\hsize=0.13\hsize}X
  >{\raggedleft\arraybackslash\hsize=0.175\hsize}X
  >{\raggedleft\arraybackslash\hsize=0.19\hsize}X}
\toprule
&Mode-classifier&Logistic Ordinal Regression&Multinomial GBT\\ 
\midrule
Accuracy, VR Validation Set&0.691&0.647&0.731\\ 
Accuracy, Labeled VRs&0.743&0.706&0.778\\  
\midrule
Weighted F1, VR Validation Set&0.565&0.473&0.682\\ 
Weighted F1, Labeled VRs&0.634&0.511&0.735\\
\midrule
MAMAE, VR Validation Set&1.2&1.08&0.877\\ 
MAMAE, Labeled VRs&1.2&0.959&0.962\\
\bottomrule 
\end{tabularx}
\end{center}
\label{tab:ordmult}
\end{table}
\subsection{Ordinal Regression Reduced to Binary Classification}
\label{sec:reduction}
\par Multinomial GBT still considers labels as discrete variables ignoring their structure. But more importantly, the semantics of explanations generated from a multinomial classifier does not match the requirements. Specifically, explanations from a multiclass classifier highlight why a property is rated with a specific rating vs. all others which could lead to scenarios where a 2-stars classification is explained through the lack of a \textit{spa wellness center}, a much higher class facility. Although this explanation is correct, it is not very useful since it is unlikely that a 2-stars property can add a \textit{spa wellness center}, and does not help to understand what makes this property better than all 1-star properties which help partners, for example, to better maintain such facilities like \textit{streaming services} or \textit{garden}. Because of this, we want to introduce information about the order of the labels. Following \cite{frank2001simple} we apply a reduction from Ordinal Regression to a set of \textit{ordered} binary classifiers: four binary classifiers are constructed where classifier $k \in \{1,2,3,4\}$ estimates $Pr(y_i>k)$, where $y_i$ is the true class of example $i$. The original training set is replicated 4 times, once for each classifier: the binary label of example $i$ with multiclass label $y_i$ in classifier $k$ is positive if $y_i > k$ and negative otherwise.
This reduction approach allows us to work with binary classifiers (which are particularly well suited for explainability), and at the same time, class-order information is kept allowing us to generate explanations with the required semantics. 
\par At inference time the authors in \cite{frank2001simple} propose an analytical method to estimate the probability of an unlabeled example belonging to each class and then outputs the class that maximizes those probabilities. This approach implicitly assumes that the base binary classifiers are calibrated. Furthermore, in order to produce consistent explanations, we want to enforce consistent labeling, which means that if a property receives a 4-stars rating, it should also receive a 3, 2 and 1-star ratings. Formally, this means that if base classifier $k \in [1, 4]$ assigns a positive label to example $i$, then all predictions by classifiers $k^{\prime}<k$ must also be positive. 
To address these two issues we propose a different inference algorithm that guarantees consistent labeling and does not rely on calibrated base classifiers. The procedure runs through all the binary classifiers in class order incrementing the number of stars until a classifier outputs a negative prediction (see Algorithm \ref{inference}). One important consequence of this algorithm is that it allows us to identify what we define as the \textit{Responsible Classifier}, which is the classifier \textit{before} the first classifier making a negative classification. This classifier encodes the information about why a property is not labeled with a lower rating. Formally, the responsible classifier is computed by a function $r$ that takes a class in $[1,5]$ and outputs a classifier index in $[1,4]$ as given by the following equation:
\begin{equation}
\label{eq:resp}
    r(c) = 
    \left\{
    \begin{array}{ll}
        1 & c<3 \\
        c-1 & c \geq 3
    \end{array}
    \right.
\end{equation}

Concretely, for a property with predicted label $\hat{y}$ we can compute explanations (why is the property classified as $\hat{y}$ vs $\hat{y}-1$, Section \ref{sec:explainability}) using classifier $r(\hat{y})$ and suggestions (what is missing to reach $\hat{y}+1$, Section \ref{sec:suggestions}) based on classifier $r(\hat{y}+1)$. The semantics of these explanations match the requirements.

\begin{algorithm}
\caption{Class Rating Procedure}\label{inference}
\begin{algorithmic}
\State $X:$ property features
\State $\theta:$ 4-dimensional vector of thresholds
\Procedure{ConsistentLabeling}{$X,\theta$}
\State $\hat{y} \gets 1 $
\Comment{All properties get 1 star}
\For {$k \in [1, 4]$}
\State $\triangleright Pr(y>k | X)$ estimated with classifier $k$ (see Section \ref{sec:modeling})
\If{$Pr(y>k | X) \geq \theta_k$ \textbf{and} $\hat{y}=k$} 
\State $\hat{y} \gets k +1 $ 
\Comment Increment stars 
\EndIf
\EndFor
\State \Return $\hat{y}$
\EndProcedure
\end{algorithmic}
\end{algorithm}

After relaxing the calibration requirement, we can use any base binary classifier. Again, considering explainability requirements, we use Logistic Regression as a baseline and found good results compared to the mode-classifier but worse than Multinomial GBT. Therefore we also used GBT as the base binary classifier in the reduction, for which scalable explanation algorithms exist (see Section \ref{sec:explainability}) and found much better results compared to Multinomial GBT in all metrics and all data sets. This suggests that the combination of non linear decision boundaries with ordinal labels is an effective technique to capture the structure of our problem. Table \ref{tab:base} summarizes these findings. 
\subsection{Monotonicity Constraints}
\label{sec:mono}
Since the property descriptions are user-generated, they tend to be noisy. One example is under-reported facilities in high classes: 5 stars villas won't list \textit{hairdryer} as an amenity because it is obvious for them to provide it. This leads to some obviously positive facilities or amenities to contribute negatively towards a higher rating (examples: barbecue and children crib). To avoid this, we introduce monotonicity constraints \cite{10.1145/568574.568577}, which enforce positive contributions in all base classifiers. This allows us to encode domain knowledge as a mechanism to make our models more robust to noise in the property descriptions. We found that by introducing these constraints the model improved all metrics in all sets (see the last column in Table \ref{tab:base}). These constraints are also crucial to produce robust and consistent explanations: the lack of a facility cannot explain why a property is rated as 3 stars as opposed to 2, monotonicity prevents this scenario. 
\par As a summary of our modeling considerations, we obtained the best performance with Ordinal Regression reduced to Binary Classification with Gradient Boosted Trees with monotonicity constraints. Equally important, this model allows us to generate robust, consistent, and scalable explanations, as described in the following sections.

\begin{table}
\caption{Ordinal Regression by reduction to Binary Classification with different base classifiers}
\small
\centering
\begin{tabularx}{\linewidth}{ 
  >{\raggedright\arraybackslash\hsize=0.49\hsize}X
  >{\raggedleft\arraybackslash\hsize=0.18\hsize}X
  >{\raggedleft\arraybackslash\hsize=0.15\hsize}X
  >{\raggedleft\arraybackslash\hsize=0.17\hsize}X}
\toprule
 & Logistic Regression& Gradient Boosting Trees & GBT + monotonicity\\ \midrule
Weighted F1, VR Validation Set &0.666 &0.729&0.732   \\ 
Weighted F1, Labeled VRs& 0.737&0.773 &0.776 \\
\midrule
MAMAE, VR Validation Set &0.839&0.633& 0.6\\ 
MAMAE Labeled VRs&0.956 &0.922 & 0.899\\
\bottomrule 
\end{tabularx}
\label{tab:base}
\end{table}

\section{Generating Explanations}
\label{sec:explainability}
\begin{figure}
\includegraphics[width=\linewidth]{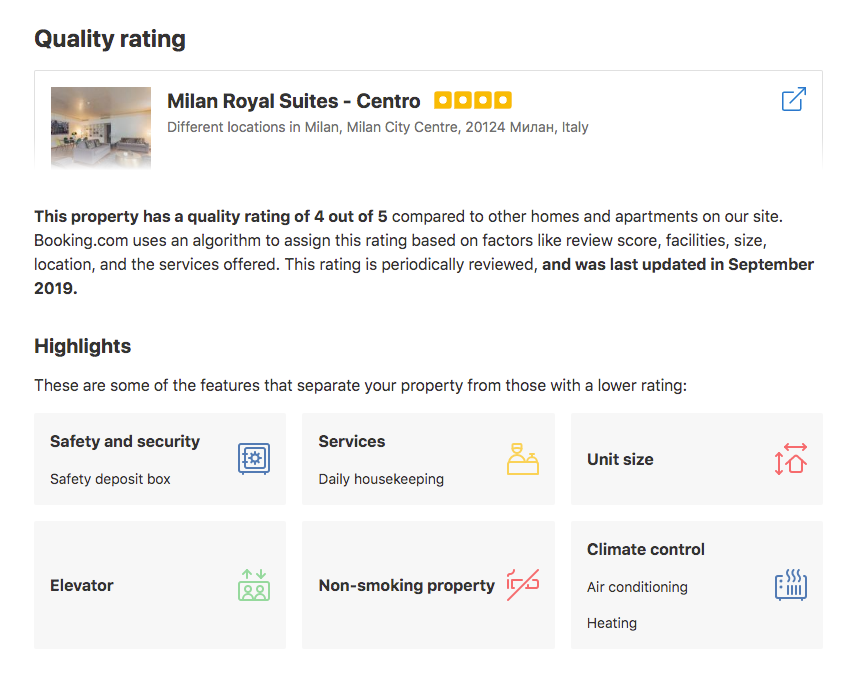}
\caption{Explaining Quality Rating in Booking.com}
\label{fig:expl}
\end{figure}
Explainability is a crucial part of our solution. We have to be able to explain to every property manager, why their property obtains a specific rating. Furthermore, as model authors, we have to be able to justify our model decisions to business stakeholders. Local and Global Interpretability play these roles respectively. According to \cite{DBLP:journals/corr/abs-1902-03501} global interpretability means \textit{"understanding the entirety of a trained model including all decision paths"}, and local interpretability, \textit{"the goal of understanding the results of a trained model on a specific input and small deviations from that input"}. Our best model is based on Gradient Boosted Trees, which is a black-box model and can't be explained directly, but since we introduced monotonicity constraints, global interpretability is possible as described in \cite{10.1145/568574.568577}. To achieve local interpretability, we applied SHAP (SHapley Additive exPlanations) by \cite{lundberg2017unified}, a framework for Machine Learning interpretability based on Shapley values \cite{shapley1953value}. In particular, TreeShap which reduced the Shapley values computation from exponential to polynomial time for tree based model \cite{lundberg2020local2global}. The semantics of the SHAP values is: given the current set of feature values, the contribution of a feature value to the difference between the actual prediction and the mean prediction is the estimated Shapley value. With this, we can provide local interpretability per property and improve global interpretability per binomial classifier using Shapley values aggregations. Since we use a reduction to several binary classifiers, we first identify the base classifier responsible for the predicted rating using Equation \ref{eq:resp} and then calculate SHAP values $\Theta$ for classifier $r(\hat{y}_i)$ (see Algorithm \ref{explanations}). The list of attributes is ranked by SHAP values and presented to the property manager of that specific property. 
Table \ref{tab:explanations} shows some examples of explanations computed with this algorithm.

 \begin{table*}
    \caption{Explanations examples, parenthesis indicates negative score}
    \centering
 \begin{tabularx}{\textwidth} { 
  >{\raggedright\arraybackslash\hsize=0.3\hsize}X 
  >{\raggedright\arraybackslash\hsize=0.7\hsize}X}
    \toprule
    Predicted class & Important features based on Shapley values   \\
    \midrule
    Class 1, $Pr(>1)=0.27$& balcony, hair dryer, garden, (size, shared bathroom, no wardrobe closet)\\
    Class 3,  $Pr(>2)=0.98$, $Pr(>3)=0.14$& dishwasher, electric kettle, cable and satellite channels, hair dryer, (non feather pillows)\\
    Class 5, $Pr(>4)=0.54$&swimming pool, daily maid service, safe deposit box, spa wellness center, (street parking)\\  \bottomrule
 \end{tabularx}
  \label{tab:explanations}
    \end{table*}

\begin{algorithm}
\caption{Explanations generation procedure}\label{explanations}
\begin{algorithmic}
\State $X$: Binary features of a property  
\State $\hat{y}:$ Assigned rating
\Procedure{ComputeExplanations}{$X$, $\hat{y}$}
\State $\triangleright$ Identify the responsible model $r(\hat{y})$ as defined by Eq. \ref{eq:resp}
\State $cl \gets r(\hat{y})$
\State $\triangleright$ internal model state required to invoke TREESHAP\_PATH 
\State $\{v, a, b, t, r, d\} \gets Unpack(cl)$
\State $\triangleright$ See Algorithm 2 in \cite{lundberg2020local2global}
\State $w \gets TREESHAP\_PATH(X, \{v, a, b, t, r, d\})$ 
\State \Return $w$
\EndProcedure
\end{algorithmic}
\end{algorithm}
    
\section{Generating Suggestions}
\label{sec:suggestions}
\begin{figure}
\includegraphics[width=\linewidth]{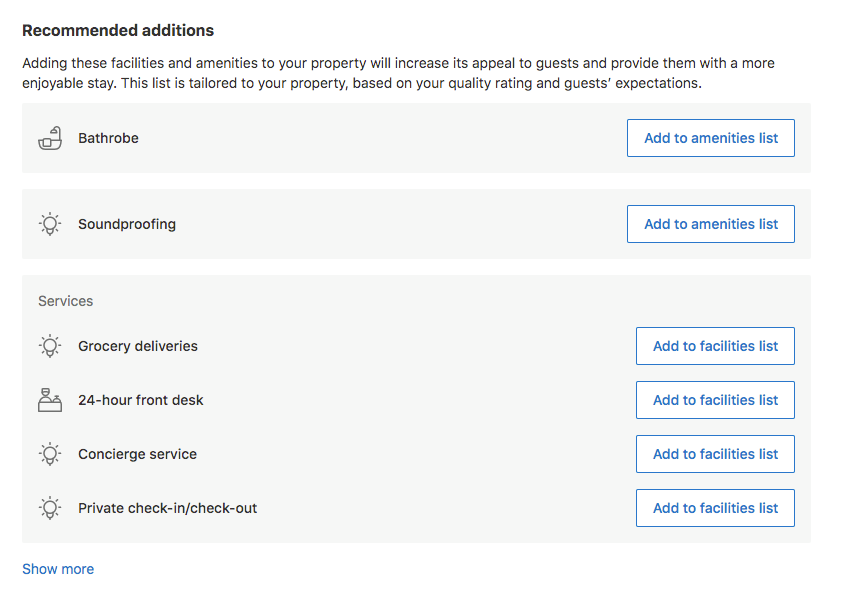}
\caption{Suggestions for a 3-stars Vacation Rental}
\label{fig:suggestions}
\end{figure}
We are also interested in explaining what is missing to reach the next rating level (e.g. add barbecue, crib, coffee machine, etc.). These explanations work as suggestions to improve the quality of a property. Two requirements must be satisfied:
\begin{enumerate}
\item  Adding the recommended feature must increase the probability of getting a higher class
\item Adding the recommended feature must not increase the probability of getting a lower class
\end{enumerate}
To this end, we only consider binary facilities (e.g. \textit{has barbecue}) and ignore non-binary ones such as \textit{the number of rooms} or \textit{size}. 
\par The procedure works as follows: for a given property with currently assigned rating $\hat{y}$, for each eligible facility with a negative value, we estimate the increment $w$ in the probability of belonging to the next class $\hat{y} + 1$ given that the corresponding feature is set to positive:
\begin{equation}
\label{eq:inc}
    w = Pr(y>\hat{y} | X^{j=1}) - Pr(y>\hat{y} | X)
\end{equation}
Where $X$ is the current feature vector containing all eligible binary features and $X^{j=1}$ is the same vector with feature $j$ flipped from $0$ to $1$. These probabilities are estimated using the responsible classifier for the next class $r(\hat{y}+1)$ when $\hat{y}<5$ and $r(\hat{y})$ when $\hat{y}=5$ (see Section \ref{sec:modeling}). Due to monotonicity constraints, such increment can only be greater than or equal to zero, guaranteeing both requirements are satisfied. Facilities are ranked by the increment in descending order and suggested to property managers. A more formal description of this procedure is presented in Algorithm \ref{suggestions}. Figure \ref{fig:suggestions} shows an example of the recommendations presented to end users.

\begin{algorithm}
\caption{Suggestion generation procedure}\label{suggestions}
\begin{algorithmic}
\State $X$: Binary features of a property  eligible for suggestions
\State $\hat{y}:$ Assigned rating
\Procedure{ComputeSuggestions}{$X$, $\hat{y}$}
\State $S \gets \emptyset$ 
\Comment{$S$: List of suggestions to be returned}
\For{$X_j \in X$}
\If{$X_j = 0$}
\State $X' \gets X$
\Comment{Copy full feature vector}
\State $X'_j \gets 1$
\Comment{Flip current feature}
\State $\triangleright$ Computed with classifier $r(\hat{y}+1)$,  see Eq. \ref{eq:resp}
\State $w \gets Pr(y>\hat{y} | X') - Pr(y>\hat{y} | X)$
\Comment{See Eq. \ref{eq:inc}}
\If{$w > 0$}
\State $S \gets S \Cup (X_j, w)$
\Comment{Add to suggestions list}
\EndIf
\EndIf
\EndFor
\State \Return $S$
\EndProcedure
\end{algorithmic}
\end{algorithm}

\section{Experiments}
\label{sec:experiments}
We validated our system by conducting Online Controlled Experiments in Booking.com, one of the largest Online Travel Platforms in the world. In our case, we have to be extra careful with the experiments since we cannot simply change the quality rating of a property with the purpose of experimentation. Therefore, the purpose of our experiments is not to benchmark different algorithms for which we rely on offline experiments but to test hypotheses about the efficacy of our system with respect to user behaviour represented by both guests and property managers. We describe four experiments that we consider relevant for this paper. All hypotheses are tested with 90\% significance level, only statistically significant results are provided together with 90\% confidence intervals on pre-registered metrics.

\begin{table}[!ht]
\caption{Experiments 1 and 2, results with 90\% CIs. All statistical significant with p-value $<$ 0.001}
\centering
\begin{tabular}{lrr}
\toprule
Metric Uplift (\%)& Experiment 1 & Experiment 2\\
\midrule
Class Filter Usage &  0.17\% ±0.05\% & 6.5\% ±0.14\% \\
CTR after Filtering by Class & 1.58\% ±0.25\% & 1.78\% ±0.45\% \\
Property Type Filter Usage & 1.05\% ±0.16\% & 4.31\% ±0.12\% \\
CTR a. Filter by Property Type & 1.43\% ±0.30\% & 0.37\% ±0.31\%\\
Rated VR CTR & 0.41\% ±0.05\% & 0.52\% ±0.05\% \\
Rated VR Conversion & 1.35\% ±0.46\% & 1.45\% ±0.54\%\\
Customer Service Tickets & No effect & No effect \\
\bottomrule
\end{tabular}
\label{tab:exps12}
\end{table}

\begin{table}[!ht]
\caption{Experiment 3: Explanations. Results with 90\% CIs. All statistical significant with p-value $<$ 0.001}
\centering
\begin{tabular}{lr}
\toprule
Metric Uplift (\%)& Experiment 3\\
\midrule
Amenities Added &  0.17\% ±0.05\%\\
Room Added  & 1.42\% ±0.51\% \\
Room Edited  & 1.21\% ±0.11\% \\
Customer Service Tickets & No Effect \\
\bottomrule
\end{tabular}
\label{tab:exps}
\end{table}

\begin{table}[!ht]
\caption{Experiment 4: Suggestions. Results with 90\% CIs. All statistical significant with p-value $<$ 0.001}
\centering
\begin{tabular}{lr}
\toprule
Metric Uplift (\%)& Experiment 4\\
\midrule
Visit Amenities & 15.86\% ±3.39\%\\
Amenities Changed  & 19.35\% ±4.75\% \\
Customer Service Tickets & No Effect \\
\bottomrule
\end{tabular}
\label{tab:suggestions}
\end{table}

\paragraph{Experiment 1} Hypothesis: "Machine Learned VR Quality Ratings help \textit{guests} to find suitable Vacation Rentals." In this experiment all eligible properties received a class rating according to the Multinomial GBT model (Section \ref{sec:baselines}). Visitors of the Booking.com website were randomly uniformly split into two groups: the control group with users exposed to the normal experience where Vacation Rentals do not have any rating (although all eligible VRs do have a label assigned, the user interface ignores them keeping everything exactly as if no ratings were available) and the treatment group where users are exposed to the machine generated quality ratings by displaying them in the search results page, but also by making them available for filtering, sorting, etc. (see Figure \ref{fig:tiles}). In the spirit of transparency, the automatically generated star-ratings are displayed with a different icon and referred to as \textit{tiles} and a generic explanation is displayed to users \textit{on-hover} stating that the tiles are indeed generated automatically.
Experiment run-time was more than 4 weeks and impacted ~100M guests and more than 500k vacation rentals. 
Results show that indeed the funnel is much more efficient since users can find more VRs (for example by filtering by class or property type) producing higher Click-through and Conversion rates. At the same time, we saw no effect on Hotels Conversion Rate, which shows that there is no cannibalization, likely because our system is improving the conversion on users that are interested only in VRs. Finally, we found no effect on Customer Service Tickets. Results are summarized in Table \ref{tab:exps12}.
\paragraph{Experiment 2} Hypothesis: "Monotonous Ordinal Regression helps guests to find suitable Vacation Rentals." In this experiment we evaluated our best performing model according to offline evaluation criteria (Section \ref{sec:mono}). It is identical to Experiment 1 but tested on a separate group of properties (no rating was changed, we only added more ratings). Run time was 2 weeks and impacted ~100M guests and more than 200k Vacation Rentals. Again, we can see the funnel improving significantly which we interpret as evidence of this model being effective. Furthermore, most effects are larger compared to Experiment 1, suggesting this Monotonous Ordinal Regression is better than Multinomial GBT. We want to remark that explanations and suggestions that meet the established requirements are only computable based on this model, therefore, we consider it superior. Experiment results are summarized in Table \ref{tab:exps12}. 
\paragraph{Experiment 3} Hypothesis: "Explanations are clear and help partners understanding how the rating was assigned." In this experiment we evaluated explanations based on Shapley Values from our best performing model according to offline evaluation criteria. In this case we split all vacation rentals for which a machine generated rating exists into two groups (as opposed to website visitors) and change the Property Management Interface. For the properties in the control group the machine generated ratings are displayed, but no explanation is available. For the properties in the treatment group, explanations are displayed as depicted in Figure \ref{fig:expl}. Run-time was 4 weeks and impacted ~200k Vacation Rentals. If the hypothesis is correct, we expect partners to improve the description of their property, otherwise, a raise in Customer Service Tickets would be observed, due to complaints about the received rating and/or explanation. Furthermore, we conducted an online survey on partners asking \textit{"Do you find the information regarding your Quality Rating and Highlights helpful?"}
We found no effect on Customer Service Tickets (by non-inferiority test), and found conclusive positive results on several metrics related to property details submission (see Table \ref{tab:exps}). The survey showed positive results with 72.3\% positive answers. From this data, we conclude that explanations are relevant and effective, supporting the hypothesis.

\paragraph{Experiment 4} Hypothesis: "Suggestions are relevant and clear, partners will visit facilities/amenities and update them." 
In this experiment, we evaluated suggestions by adding the "Recommended additions" section (described in Figure \ref{fig:suggestions}) to the property management page. The setup is the same as Experiment 3, run-time was 4 weeks and impacted ~30k vacation rentals (random sample from all eligible properties). Similarly to the previous experiment, if the hypothesis is incorrect, we expect a raise in Customer Service Tickets due to complaints about the irrelevant suggestions. If the hypothesis is correct, we expected partners to visit facilities/amenities sections of their property more often and more changes in the facilities and amenities.
We found no effect on Customer Service Tickets (by non-inferiority test), and found conclusive positive results on several metrics related to property details submission (see Table \ref{tab:suggestions}). From these results, we conclude that the suggestions are indeed helping partners to improve their properties or their descriptions.

\section{Conclusion}
\label{sec:conclusion}
\begin{figure*}
\includegraphics[width=0.7\linewidth]{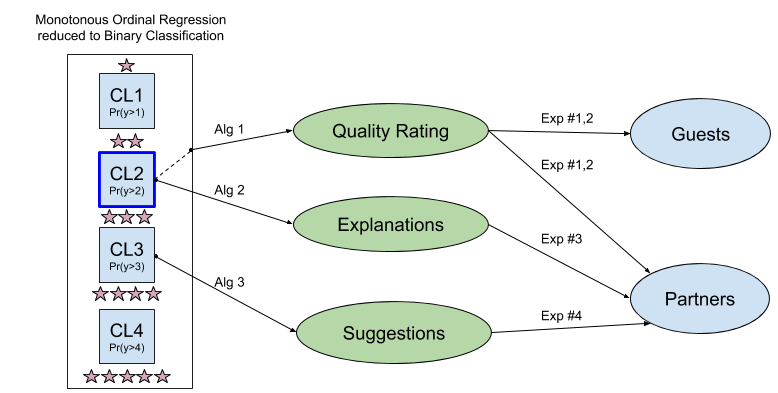}
\caption{From models to Users. Highlighted classifier is the responsible classifier described in Section \ref{sec:reduction}}.
\label{fig:comp}
\end{figure*}

In this work we presented a Quality Rating System for Vacation Rentals based on Machine Learning. Several challenges were addressed, and technical details discussed, with rationale about design choices and trade offs. Our solution hinges on 3 main points: Collaborative Labeling that allowed us to transfer hotels Star Ratings to Vacation Rentals, Ordinal Regression by reduction to Binary GBTs with Monotonicity constraints which successfully captures the order in the classes allowing both accurate and explainable predictions with correct semantics, and SHAP that allowed to compute consistent explanations. The effectiveness of the system was thoroughly validated through massive Online Controlled Experiments conducted in Booking.com, one of the top Online Travel Platforms in the world with millions of daily users and millions of Vacation Rentals, showing strong evidence of significant benefits for all the relevant parties: 
\begin{itemize}
\item Guests: it is easier for them to find accommodation fitting their needs and preferences.
\item Partners: they improve their commercial health through better market targeting, better visibility and through insights about how to improve and maintain the provided quality.
\item Online Travel Platform: it benefits from more engaged users and partners and more transactions.
\end{itemize}

Figure \ref{fig:comp} illustrates the relationship between the applied techniques and the parties involved in the platform.
\par We believe that integrating user facing explanations into the modeling process was fundamental to find a good balance between accuracy and explainability. How to improve this trade-off, maybe through white-box models, is an interesting and promising direction for future work.

\balance
\begin{acks}
This work is the result of the contributions of a large group of professionals. Authors want to thank Ioannis Kangas for sharing ideas and initial prototypes around Collaborative Labeling, Thomas Bosman for his significant contributions to the Ordinal Regression Reduction, Ahmed Khaili for his contributions around Feature Generation, Roberto Pagano and Yan Romanikhin who collaborated with the implementation and Dennis Bohle for his insights about explainability.
\end{acks}

\bibliographystyle{ACM-Reference-Format}
\bibliography{miningthestars} 

\end{document}